# Foveated Model Observers for Visual Search in 3D Medical Images

Miguel A. Lago, Craig K. Abbey, and Miguel P. Eckstein

***Abstract*** — Model observers have a long history of success in predicting human observer performance in clinically-relevant detection tasks. New 3D image modalities provide more signal information but vastly increase the search space to be scrutinized. Here, we compared standard linear model observers (ideal observers, non-pre-whitening matched filter with eye filter, and various versions of Channelized Hotelling models) to human performance searching in 3D $1/f^{2.8}$ filtered noise images and assessed its relationship to the more traditional location known exactly detection tasks and 2D search. We investigated two different signal types that vary in their detectability away from the point of fixation (visual periphery). We show that the influence of 3D search on human performance interacts with the signal's detectability in the visual periphery. Detection performance for signals difficult to detect in the visual periphery deteriorates greatly in 3D search but not in 3D location known exactly and 2D search. Standard model observers do not predict the interaction between 3D search and signal type. A proposed extension of the Channelized Hotelling model (foveated search model) that processes the image with reduced spatial detail away from the point of fixation, explores the image through eye movements, and scrolls across slices can successfully predict the interaction observed in humans and also the types of errors in 3D search. Together, the findings highlight the need for foveated model observers for image quality evaluation with 3D search.

## I. INTRODUCTION

Model observers are mathematical formulations that can be applied to image statistics and actual images to predict human observer performance in simple visual tasks such as detection and classification of luminance-defined signals [1]–[5]. Through four decades, the field of medical imaging has made progress in the development and evaluation of model observers. Work in the 1980s and early 1990s focused on computer-generated backgrounds and simple detection tasks with a single signal appearing at one or a few locations (location-known exactly, LKE) [6]–[12]. Since then, model observers have been extended in several important ways to try to capture the complexities of clinical perceptual tasks with medical images. Since the mid-1990s, model observers were first applied to large samples of anatomical backgrounds from digitized medical images [13]–[20], more complex classification tasks [21], signals that could vary in size and shape [6], [18], [20], [22], and dynamic or 3D components [23]–[31].

Still, a majority of studies have evaluated the validity of model observers with simple detection tasks for which the signal might appear in a single or few known locations within the image [6], [18], [22], [28], [32]–[44]. The underlying assumption is that model observer performance for simple visual detection tasks with a reduced number of possible locations reliably reflects performance in more clinically realistic tasks, which involve searching for the abnormality across a larger area in the image. If so, conclusions from evaluations and optimizations of imaging systems with simple LKE tasks will be applicable and valid in more clinically realistic search tasks.

Studies have used methods based on signal detection theory to quantitatively relate performance across a varying number of possible locations and even free search over an entire 2D image [8], [13], [45]–[47]. However, in recent years, researchers have highlighted the limitations of the simple LKE detection tasks and motivated the need to develop model observers that account for the process of visual search [48]–[52]. Furthermore, a recent study has highlighted discrepancies between the rank ordering of imaging systems based on location-known detection and search tasks [53].

Also, the last decade has seen an increase of 3D imaging techniques consisting of multiple slices that are quickly replacing traditional 2D techniques (digital breast tomosynthesis [53], computed tomography [54], 3D ultrasound [56], and magnetic resonance imaging [57]). These 3D images provide more information about the signal and reduce tissue overlap compared with traditional 2D imaging modalities like projection x-ray imaging or nuclear scintigraphy. On the other hand, the increase of information from the volumetric data comes with the caveat of more uncertainty in the possible location of a signal, which is well-known to degrade model and human performance [8], [13], [45], [58], [59].

During visual search, humans make approximately three eye movements per second [60], [61] to point their fovea at objects of interest. The fovea is the central region of the visual field, where acuity is the highest. Extra-foveal visual processing, also referred to as peripheral processing, is mediated by a reduced density of retinal photoreceptors (cones) [62], higher spatial integration of cone outputs onto retinal ganglion cells, and fewer neurons in the primary visual cortex per millimeter of the retina [63]. As a consequence, the visual periphery lacks the capability for fine spatial discriminations. Visual forms surrounding the signal will also have additional adverse effects

This work was supported by the National Institute of Health grant R01EB026427.
M. A. Lago is with the University of California, Santa Barbara, Santa Barbara, CA 93105 USA (e-mail: lago@psych.ucsb.edu).
C. K. Abbey is with the University of California, Santa Barbara, Santa Barbara, CA 93105 USA (e-mail: abbey@psych.ucsb.edu)
M. P. Eckstein is with the University of California, Santa Barbara, Santa Barbara, CA 93105 USA (e-mail: eckstein@psych.ucsb.edu)

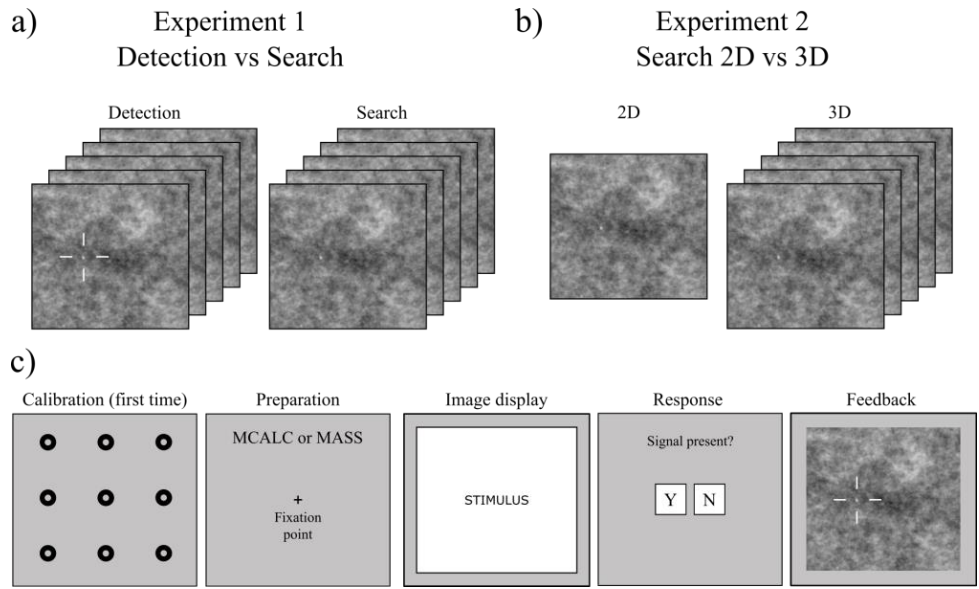

Figure 1: a) Sample of a stimulus for Location Known Exactly task (LKE) for 2D and 3D conditions. b) Sample of a stimulus for search task for 2D and 3D conditions. c) Timeline for a single trial of the experiments. Experiment starts with an eye-tracker calibration prior to starting the trials. On each trial, a preparation screen is shown, indicating the signal type to be searched. The image is presented (2D or 3D), and the participant is given unlimited time to search. When the participant decides to terminate the search, the response screen is shown. Finally, if the signal was present, a feedback image with the correct location of the signal is presented.

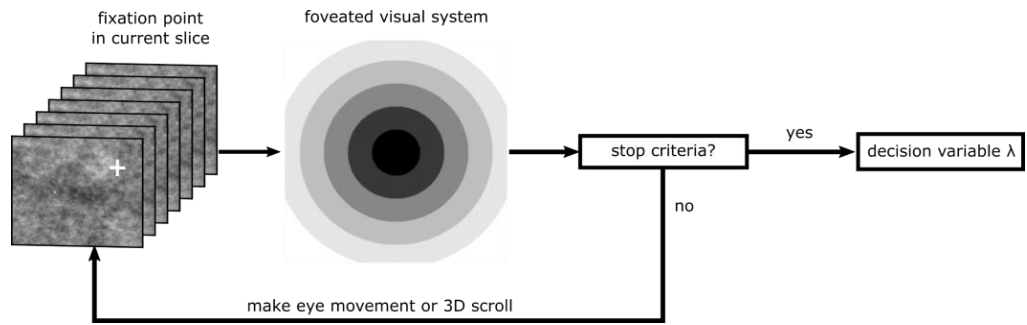

Figure 2: Schematic for the foveated search model (FSM). Given a specific fixation, the FSM processes the current slice at a specific fixation point with different templates at different eccentricities capturing the properties of the foveated human visual system. The templates are an optimal linear combination of different sets of eccentricity-specific Gabor channels. Subsequently, the FSM compares the maximum likelihood ratio for signal-presence evidence and compares it with a threshold. If the likelihood ratio surpasses this threshold, the search stops, and the model outputs a decision variable; otherwise, it makes another eye movement or 3D scroll. At the end of search the model takes a maximum across slices and compares the decision variable to a decision threshold to make yes/no responses.

on signal detectability at the visual periphery [64], [65]. Therefore, extra-foveal processing can reduce the detectability of certain types of signals in medical images [66], [67].

For a detection task with few known locations, humans will typically scrutinize all locations by moving their eyes and pointing their central vision (the fovea) to the regions possibly containing the signal. For 2D images, the process is more laborious depending on the image size, but radiologists still likely foveate a large percentage of each image through eye movements. With 3D modalities, the most common method to read images in radiology is by viewing one 2D section or a slice of the volume at a time and freely scrolling through the slices [68], [69]. Radiologists face the challenge of searching through the "stack" of slices in a short amount of time [70]. However, unlike 2D images, given the amount of information that 3D images have, a human reader is unlikely to explore each region of each slice foveally. Instead, readers may process much of the information with peripheral vision [67] and rely on guided eye movements to fixate suspicious signal locations. If observers rely on peripheral visual processing, search performance can be significantly impacted. The influence will depend on the detectability of the signal in the visual periphery and the eye movements' effectiveness in guiding the fovea to the signal location. The increased reliance on peripheral processing with 3D images might lead to performance differences from those obtained from LKE. Also, current model observers in the literature incorporate properties of the human visual system only at the fovea [11], [12], [33], [37], [38], [71], [72] but do not account for observer's vision at the visual periphery. Thus, existing model observers might be unable to successfully predict search performance in 3D images for signals which detectability varies greatly from the fovea to the visual periphery.

While there is a growing literature on the interactions of search strategies and the human perceptual system in 2D images [61], [73]–[78] and more recently 3D [23], [67], [79], [80], few studies have focused on visual search in 3D images and the role of peripheral visual processing [67], [81].

In this context, the first goal of the current paper was to demonstrate experimentally how visual search in 3D images introduces a new component to human observer performance which can greatly influence signal detectability and might create dissociations with LKE detection and search performance in 2D images. This new component of 3D search relates to the visual processing of the image data with extra-foveal regions of the retina. The paper's second goal was to evaluate whether existing model observers in the literature, not accounting for the process of search and vision away from the fovea, could predict human performance in 3D visual search. The third goal was to propose a new model observer that incorporates processing with variable resolution across the visual field (foveated vision), eye movements, and 3D scrolls (foveated search model, FSM) and to evaluate its ability to predict variations of human performance across LKE tasks, 2D and 3D search.

To assess the task's influence on 3D images, we first studied human performance in a location known exactly 3D detection task and 3D search (experiment 1). We then compared human performance in 2D search vs. 3D search (experiment 2). We expected the effects of 3D visual search to interact with the signal's detectability in the visual periphery and lead to dissociations with 3D LKE detection and 2D search. Thus, we evaluated human performance with two different signals with different detectabilities in the visual periphery. We evaluated five different existing model observers: non-prewhitening matched filter with an eye filter, Hotelling observer/ideal observer, and channelized Hotelling observer with Gabor channels, Difference of Gaussian (DoG) channels, and Laguerre Gauss channels. Finally, we proposed a model that processes the entire visual field with varying spatial resolutions (foveated search model, FSM), explores the stimulus through eye movements and scrolls, and makes decisions by integrating visual information across the search process. We ran a supplementary experiment to measure signal detectability as a function of retinal eccentricity while observers maintained their gaze on a fixation point. This third experiment was used to fit the parameters of the FSM's foveated component to match human deterioration of signal detectability with increasing eccentricity.

## II. MATERIALS AND METHODS

### A. Display and Image generation

A medical-grade monitor of 1280×1024 resolution (Barco MDRC-1119 LCD monitor) at a 75cm distance from participants' eyes was used to display these stimuli. We calibrated the monitor linearly between 0.1 cd/m$^2$ and 111 cd/m$^2$ for gray levels 0 and 255, respectively. The stimuli were generated as 3D correlated noise backgrounds by filtering white noise ($\mu = 128, \sigma = 25$) by a power spectrum of $1/f^{2.8}$ using frequency indexes. The image size was 1024×820×100 voxels. We generated two 3D signals that were embedded in the already generated backgrounds (Supplementary Figure 1). Signals were linearly added with no filtering to keep their frequency-space properties. The first signal resembled a small microcalcification as a sharp spherical sphere of 6 voxels diameter (~0.13 degrees of visual angle), constant contrast. The second signal was a 3D Gaussian blob ($\sigma$ = 10 voxels, ~0.55 degrees of visual angle), considered to be closer to a larger mass. One of the two signals (50% chance) was presented in a random location in 50% of the trials. Additionally, a 2D version of these images was generated by selecting a single slice of the volume (the signal's central slice, if present). All experiments used the same backgrounds. Signal contrasts varied across experiments because of the varying difficulties of the LKE and search tasks. The signal contrasts were selected to avoid ceiling effects (100 % accuracy) and floor effects (chance performance) based on pilot experiments. Contrast was defined as the peak signal amplitude divided by the mean background amplitude. The signal contrast of the two signals was kept constant across conditions for each experiment.

Table 1: Summary of experiments. * = same participants did these two studies.

|  | LKE Det. vs Search | 3D vs 2D Search | Forced Fixation |
|---|---|---|---|
| # participants | 6 | 7* | 7* |
| # trials | 300 | 800 | 4,000 |
| Signal peak contrast | 0.45 | 0.65 | 0.65 |

### B. Psychophysics Experiment 1: Location Known Exactly (LKE) 3D detection vs. 3D search.

Observers were instructed to detect or search through a 3D image stack for one of the two signals that varied their position and slice across trials when present (50 % probability). A high contrast copy of the signal was presented above the stimuli for reference. For the LKE detection task (but not the 3D search), a fiduciary cue marked the signal's potential location, and the initial slice was the central slice of the signal (if present). For search, the initial slice was the first slice of the volume. Observers scrolled freely using the mouse and a non-overlapping scroll bar drawn on the right side of the screen. Observers had no time limit to explore and pressed the spacebar to end the trial and respond about the presence of the signal (yes/no). Subsequently, feedback was provided about the correctness of the response and signal location for signal-present trials. We used an infrared video-based eye tracker (Eyelink 1000, SR Research Inc.) to detect saccades (eye velocity and acceleration thresholds of 30 degrees/sec and 9,500 degrees/sec$^{2,}$ respectively). Scrolls were also recorded at the sampling rate of the screen (60Hz). Signal peak contrast was set to 0.45 for both signals to avoid ceiling or flooring effects. Figure 1a shows an example of the stimuli used for this experiment. Figure 1c shows the timeline for one trial.

### C. Psychophysics Experiment 2: 2D vs. 3D search

The search experiment in 2D and 3D had no fiduciary marks. For the 3D search condition, the observers were presented first with the first slice of the stack. Given that the search task is harder, the signal peak contrast in this case was set to 0.65 for both 2D and 3D and both signals. Figure 1b and Figure 1c show the stimuli used for this experiment and the timeline for one trial, respectively.

*D. Psychophysics Experiment 3: Forced Fixation*

A third experiment was designed to measure how the signal detectability decreased as a function of retinal eccentricity, E (in degrees of visual angle). This data is required to fit the foveated model's parameters to match the peripheral degradation of signal detectability measured in human observers. The experiment used the same 2D stimuli and signals as described in subsection II.A. Participants were asked to fixate at a cross while fiduciary cues indicated the potential location of the signal. We refer to this task as a forced fixation task. To start the trial, observers pressed the space bar while fixating at a given fixation location, and the stimulus was presented for a brief time (500ms). Participants responded as to whether the signal was present (yes/no) at the cued location. If the eye tracker detected a saccade, the trial was interrupted. Feedback was given after every trial about whether their decision was correct. To compare results with the search task, signal peak contrast was also set to 0.65.

*E. Participants*

Six observers participated in the 3D detection vs. 3D search experiment for 150 trials in each condition. A different group of seven observers participated in both the search (400 trials for 2D and 400 for 3D) and the forced fixation experiment (4,000 trials). All participants were undergraduate students from the University of California, Santa Barbara, who had verified normal or corrected-to-normal vision. Gender balance was 28% male and 72% female, and ages ranged from 20 to 23 years. They were provided informed consent and treated according to the approved human subject research protocols by the University of California, Santa Barbara: 12-18-0025, 12-16-0806, and 12-15-0796.

*F. Model Observers*

We calculated the performance of three traditional non-foveated model observers in both 2D and 3D: the Ideal Observer (Hotelling), the Non-Prewhitening Model Observer with Eye Filter (NPWE), and the Channelized Hotelling Observer (CHO). The model observers were implemented as scanning models that compute a decision variable for all locations in the image [48], [50]–[52], [79]. Here, we describe the 3D implementation of these models. The 2D templates can be derived similarly, as described in the literature [2], [4], [6], [9], [82]. The template, **w**, for each observer was convolved ($*$) with the image, **g**. The response was computed as:

$$\boldsymbol{\lambda} = \mathbf{w} * \mathbf{g} + \boldsymbol{\epsilon}_{\text{int}}, \quad (1)$$

where $\boldsymbol{\lambda}$ is an array of responses at each individual voxel, $\boldsymbol{\lambda} = [\lambda_v \ldots]$, $v = 1, \ldots, V$, and $\boldsymbol{\epsilon}_{\text{int}}$ is the internal noise. These scalar responses were combined to calculate the likelihood of the absence and presence of the signal:

$$\mathcal{L}_v = \frac{1}{\sigma\sqrt{2\pi}} \exp\left(\frac{-(\lambda_v - \mu)^2}{2\sigma^2}\right) \quad (2)$$

where $\mu$ and $\sigma$ are the mean response and standard deviation of the template response, respectively. The likelihood ratio ($\mathcal{LR}$) of the signal-present and the signal-absent likelihoods was used as the decision variable:

$$\mathcal{LR}_v = \frac{\mathcal{L}_v^+}{\mathcal{L}_v^-} \quad (3)$$

For the ideal observer, a yes/no decision was based on an optimal sum of likelihoods across locations ($\sum \mathcal{LR}_v$), while the other model observers used the maximum response ($\max(\mathcal{LR}_v)$). These models did not incorporate any internal noise ($\boldsymbol{\epsilon}_{\text{int}} = 0$). Below, we describe the templates (i.e., **w**) for each model.

*1) 3D Ideal Observer*

The template for the ideal observer (IO) is calculated from the 3D image covariance matrix of the background noise $\mathbf{K}_g$ combined with a vector representing signal **s** as follows:

$$\mathbf{w}^{\text{IO}} = \mathbf{K}_g^{-1}\mathbf{s} \quad (4)$$

*2) 3D NPW Observer*

The template for the non-prewhitening matched filter observer coincides with the signal profile:

$$\mathbf{w}^{\text{NPW}} = \mathbf{s} \quad (5)$$

*3) 3D NPWE Observer*

To generate the template for the non-prewhitening observer with eye filter (NPWE) we used the mathematical representation of the signal and a 2D contrast sensitivity function as an eye filter [82]. To create the 3D template, we applied the same spatial eye filter to the template in different slices and combined them in one 3D template. The spatial eye filter was generated as follows:

$$\xi(\rho) = \rho^\alpha \exp(-\beta\rho^\gamma) \quad (6)$$

using the parameters by Bochud et al. [41] ($\alpha$=1.4, $\beta$=0.013, and $\gamma$=2.6) where $\rho = \sqrt{u^2 + v^2}$ is the radial spatial frequency in cycles per degree. The template is then calculated for each slice n of the signal as follows:

$$\mathbf{w}_n = \mathcal{F}^{-1}(\xi^2 \mathcal{F}(\mathbf{s}_n)) \quad (7)$$

where $\mathcal{F}$ stands for the Fourier transform operation.

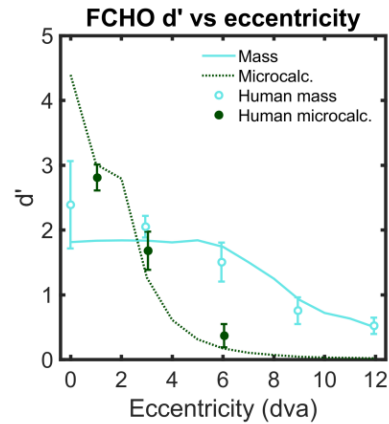

Figure 3: Human signal detectability (d') as a function of distance from the fixation point (eccentricity in degrees of visual angle, dva) and predicted performance for the FCHO model. Error bars are standard error of the mean across observers with a confidence interval of 1 standard deviation.

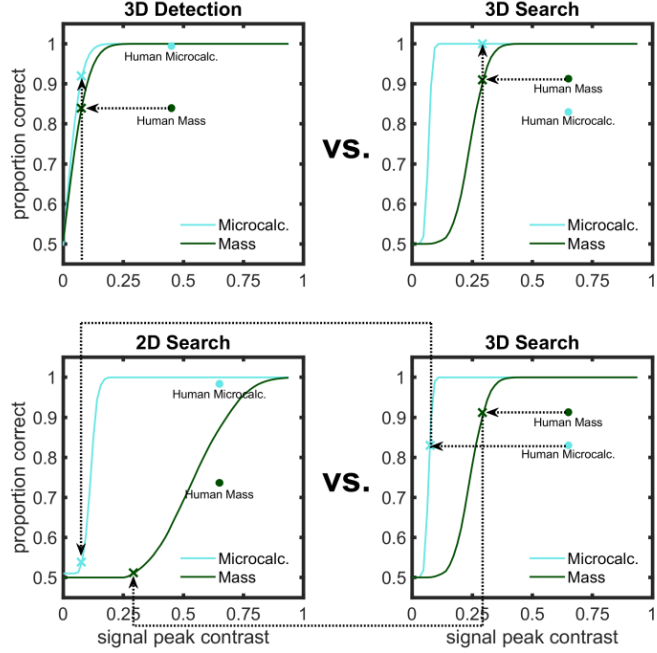

Figure 4: Proportion Correct (PC) as a function of signal peak contrasts for the ideal observer. Illustration of the process to facilitate comparisons of human and model observer performance and avoid ceiling effects. The figure illustrates the procedure for the ideal observer. A similar approach was used for other model observers except for the foveated search model (see Results). Top graphs illustrate the process for the 3D detection vs. 3D search experiment, where we were interested in the relative human/model accuracy across signals. For each task, we chose the ideal observer contrast that matched human performance for the mass (see horizontal dotted lines) and used the same contrast for the microcalcification. The bottom graphs show the procedure for the 2D search vs. 3D search comparisons. Here, we were interested in the effects of 3D search relative to 2D search for each signal. For each signal, we selected the ideal observer's contrast that matched human performance in 3D search (see horizontal dotted lines) and used that same contrast for 2D search (see arrows).

Finally, we combine the templates for all the N slices in a 3D template:

$$\mathbf{w}^{\text{NPWE}} = [\mathbf{w}_1, \mathbf{w}_2, \dots, \mathbf{w}_N] \tag{8}$$

Once the template is constructed, Eq. (1) can be used to generate responses for performance analysis.

*4) 3D Channelized Hotelling Observer*

We used a hybrid channelized Hotelling observer (CHO) based on Yu et al. [32], which computes the response of each 2D channel to each slice of the signal and combines them in one 3D template. Three different versions of this model were generated: using Gabor channels [15], Laguerre-Gauss channels [14], and Difference of Gaussian (DOG) channels [83]. Details about the channel parameters are also included in supplementary methods.

We denote the channel matrix by $\mathbf{T}$ (size: number of channels times number of pixels) and its transpose $\mathbf{T}^t$. The 2D template for each slice n is built as follows:

$$\begin{aligned} \mathbf{v}_n &= \mathbf{T}^t \mathbf{s}_n \\ \mathbf{K}_{\mathbf{v}_n} &= \mathbf{T}^t \mathbf{K}_g \mathbf{T} \\ \mathbf{w}_n &= \mathbf{T} \mathbf{K}_{\mathbf{v}_n}^{-1} \mathbf{v}_n \end{aligned} \tag{9}$$

where $\mathbf{v}_n$ are the responses of the channels to the signal, $\mathbf{K}_g$ is the covariance matrix of the background noise, and $\mathbf{K}_{\mathbf{v}_n}$ is the channel's covariance matrix. As in Eq. (8), the 2D templates are combined into a 3D template:

$$\mathbf{w}^{\text{CHO}} = [\mathbf{w}_1, \mathbf{w}_2, \dots, \mathbf{w}_N]. \tag{10}$$

### G. 3D Foveated Search Model (FSM)

This paper's main contribution is the introduction of a model that processes the visual field with varying spatial detail (foveated visual system) and explicitly fixates at locations coming from either human subjects or a fixation generation algorithm. We extended the CHO model observer and generated different model templates as a function of the distance of the signal location from the model's fixation point (retinal eccentricity). This Foveated Search Model (FSM) processes the entire image in parallel [74], [75], [84]. The FSM has several components: a foveated channelized hotelling observer model (FCHO) for detection, an eye movement model that directs the fixations, and an integration component that synthesizes information at multiple locations into a single search response. Figure 2 shows the flowchart of the model from the input image to the final decision.

*1) Foveated Channelized Model Observer (FCHO)*

The underlying model observer that the FSM uses is a modified channelized Hotelling observer with a foveation component (FCHO). To implement decreased spatial resolution at more eccentric locations, we decreased the center frequencies of the supporting Gabor channels (while maintaining their 1-octave bandwidth) in Eq. (9). At the fovea (0 degrees of eccentricity), we use the original Gabor channels from the CHO standard model: 6 spatial frequencies (32, 16, 8, 4, 2, and 1 cycles per degree) and 8 orientations in even phase [19], [40], [71]. Then, as the eccentricity increases, the central frequencies for all the Gabor channels are non-linearly scaled with respect to the eccentricity in degrees (E):

$$scaling = 1 + \alpha E^\beta. \quad (11)$$

*2) Foveated Templates that match human signal detectability vs. Retinal eccentricity*

The templates resulting from the linear combination of the channels lose access to high spatial frequencies as the eccentricity increases. We constructed 10 sets of channels to cover the image from 0 to 9 degrees of visual angle and corresponding templates (equations 8-10). We then calculated a d', signal detectability, for each template at each eccentricity. We selected the scaling parameters ($\alpha$ and $\beta$ for the FCHO, in Eq. 11) of the model to match the deterioration in human signal detectability with retinal eccentricity. We optimized these parameters to simultaneously fit the observed human d' at each eccentricity (Figure 3) for microcalcification and mass signals using the maximum likelihood method. The model also included additive internal noise sampled from a Gaussian distribution, $\epsilon_{int} \sim \mathcal{N}(0, (K\sigma_\lambda)^2)$, where the magnitude of the internal noise was proportional to the standard deviation of the template response $\sigma_\lambda$, and adds one fitting parameter K to the model. The optimized parameters were found to be $\alpha$=0.7063, $\beta$=1.6953, and K=2.7813. Figure 3 shows the human signal detectability index d' [85] as a function of retinal eccentricity and the model predictions with the best-fit parameters.

*3) Processing individual fixations*

Unlike standard model observers, the foveated model processes the entire image with distinct templates that take into consideration the distance of the image region from the point of fixation. These fixations may come from different sources. In this paper, we consider human fixations captured by the eye-tracker and a maximum a posteriori (MAP) eye-movement algorithm [75], [77], [86], [87] detailed in section 7. The FSM model also accumulates information across fixations in each 2D slice of the 3D volume. The template response at eccentricity $E$ is calculated by using a template corresponding to the retinal eccentricity as determined by the distance of the image subregion, $g$, from the current fixation:

$$\boldsymbol{\lambda}_E = \mathbf{w}_E * \mathbf{g} + \boldsymbol{\epsilon}_E \quad (12)$$

where $\boldsymbol{\epsilon}_E = [\epsilon_v, ...]$ for voxel locations at eccentricity E.

*4) Integration across Fixations*

For both human eye-movements/scrolls and MAP eye-movements/scrolls, we assumed statistical independence to integrate information across fixations. To obtain a final decision variable $\Lambda_v$ at each voxel $v$ of the current slice $z$, the output of Eq. 12 is used as the input for the likelihood ratio (Eq. 2 and 3) for all fixations ($N$) in the current slice. To integrate across fixations in the current slice, we calculated the product of their likelihood ratios:

$$\Lambda_{v,z} = \prod_{n=1}^{N} \mathcal{LR}_{v,n}. \quad (13)$$

*5) Perceptual Decision*

The perceptual decision consisted of a simple maximum across all slices and fixations,

$$\Lambda^{FSM} = \max(\Lambda_{v,z}), \quad (14)$$

which was then compared to a decision threshold to make a yes/no decision. We chose the decision threshold that maximizes proportion correct which we will refer to as the optimal threshold. Performance is calculated as the proportion of correct responses (PC).

*6) Human Eye Movement and Scrolling Model*

To guide the fixations of the foveated search model (FSM) for a given image or image stack, we used the measured sequence of human fixations and scrolls from the human observers for that same image/s. The FSM's search was terminated with the last measured human fixation of the list. We excluded short fixations (<50 ms) considered to be a result of saccade under or overshooting and followed by corrective saccades [88]. These outliers only amounted to 0.3% of all fixations.

*7) Maximum A Posteriori Eye Movement and Scrolling Model*

As an alternative to using human observer data as the model's fixations, a Maximum A Posteriori (MAP) eye movement algorithm was also used to drive the foveated search model's fixations [75], [77], [86], [87]. Given a point of fixation (starting at the center of the image), the foveated model computes the decision variable $\lambda_v$ for each voxel of the image $v$ and then infers the next fixation point. The model moves its fovea to the location with the strongest evidence of the presence of the signal:

$$f = \text{argmax}(\lambda_v). \quad (15)$$

Inhibition of return [87] was accounted for by truncating the responses already processed by the fovea to the median likelihood ratio before calculating the next fixation point. This way, we ensured the algorithm was not trapped in previously foveated areas.

Since the distribution of $\Lambda_{v,z}$ changes with the number of fixations $N$ executed by the model in a single slice, we used an optimal (PC maximizing) decision threshold specific to the number of fixations executed on a slice. We estimated the optimal thresholds for $N = [0 ... 30]$ by training the model on an independent set of trials.

To terminate the MAP search the model stopped if either of two different stopping criteria were met: 1) if the maximum $\lambda_v$ exceeded the optimal decision threshold for the current number of fixations, 2) when the explored volume (with a 2.5 degrees radius useful field of view) exceeded a threshold of 25% for microcalcification and 10% for the mass in 3D, 80% and 60% respectively for 2D. These volume thresholds were based on average explored volumes measured for the human observers.

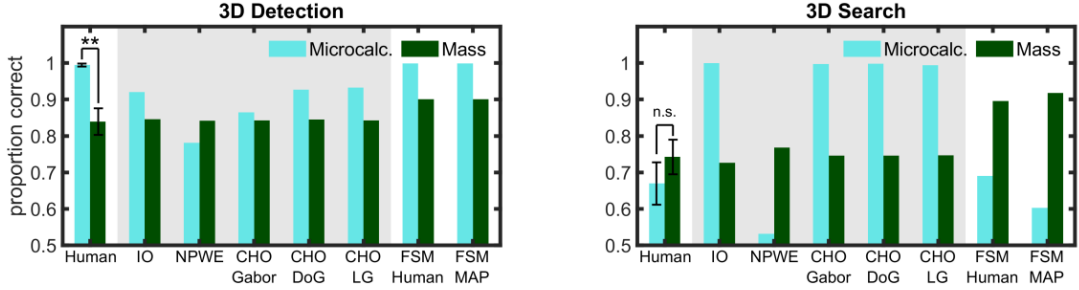

Figure 5: 3D detection vs. 3D search results. Proportion correct for human observers, ideal observer (IO), non-prewhitening with eye filter (NPWE), channelized Hotelling observer (CHO) using Gabor, DOG, and Laguerre-Gauss channels and foveated search model (FSM) for detection (left) and search (right) tasks. Model observers inside the gray area had their signal contrast reduced to match mass human performance for each condition (3D detection or 3D search). Error bars are standard error of the mean across observers, with a confidence interval of 1 standard deviation. Significance is only indicated for human observers' results. ** = p < 0.01.

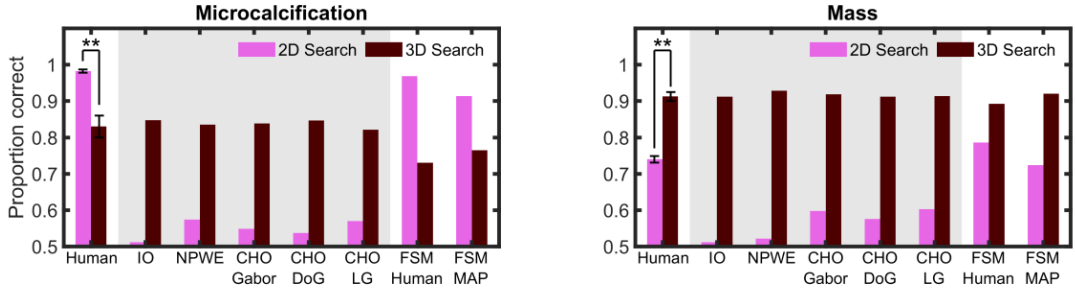

Figure 6: 2D search vs. 3D search results. Proportion correct for human observers, ideal observer (IO), non-prewhitening with eye filter (NPWE), channelized Hotelling observer (CHO) using Gabor, Difference of Gaussian, and Laguerre-Gauss channels and foveated search model (FSM) with human eye movements and MAP eye movement model for microcalcification (left) and mass (right). Model observers inside the gray area had their signal contrast reduced to match 3D human performance. Error bars are standard error of the mean across observers, with a confidence interval of 1 standard deviation. Significance is only indicated for human observers' results. ** = p < 0.01.

Another key component of the foveated search model in 3D is the scroll model. The model decides whether to move to the next slice or to make a new fixation. We considered a hybrid strategy where the model, by default, uses a drilling strategy and scrolls through the volume unless the decision variable $\lambda_v$ exceeded a threshold, in which case an eye movement is executed using the MAP algorithm. This threshold was optimized to obtain a number of scrolled slices per fixation near that of the average human data. It was set as a fraction of the optimal decision threshold at the fovea for the MAP algorithm (0.35 and 0.7 times for the microcalcification and the mass signal, respectively). Additionally, when the first or last slice is reached, the scrolling direction is reversed.

*8) 3D slice position uncertainty of the signal*

Whereas for traditional non-foveated model observers the entire image is processed by the template, the foveated search model only samples a subset of locations with the foveal template. Additionally, in 3D search, only a subset of slices might be processed. This may cause the model, on some trials, to miss the central slice of the signal and only explore slices that partially contain the signal and not the central slice. This causes a mismatch between the image data and the template and can result in lower likelihood ratio values and a consequent signal miss by the model.

To overcome this limitation, our model adds evaluations by calculating template responses for various shifts in depth for the signal. The foveated search model (FSM) considers $\delta = -2, \dots, 2$ additional 3D shifts of the five slices of its 3D template (one and two slices forward and backward from the current slice). The model then uses the maximum of the template responses $\lambda_{v,\delta}$,

$$\lambda_{v,\delta} = \mathbf{w}^{\text{FSM}} * \mathbf{g}_\delta + \epsilon_{\text{int}} \quad (12)$$
$$\lambda_v = \max(\lambda_{v,\delta}).$$

This modification allows the model to detect the signal when fixating on non-central slices of the signal of the 3D image stack.

III. RESULTS

*A. Comparison of Ideal, Human, and Model Observers Absolute Performance*

We used the statistical efficiency [10], [31] defined as the squared ratio of ideal observer and human or model peak signal contrasts (to achieve measured human performance) to assess absolute performances. Average human efficiencies ranged from 0.37 to 0.83. Model efficiencies ranged from 0.48 to 0.96.

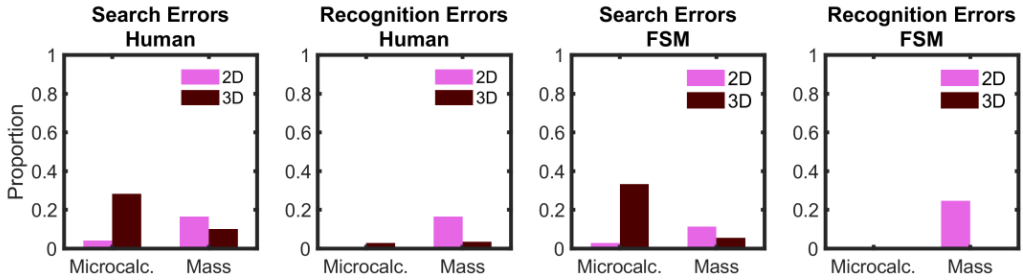

Figure 7: Search and recognition errors for false negative trials (misses) for human observers and foveated search model (using human eye movements) for the 2D and 3D search experiment. Search errors are defined as trials in which the observer did not foveate the signal and decided "signal absent" (with a tolerance of 1 degree of visual angle). Recognition errors are defined as trials in which the observer foveated the signal but after the search decided "signal absent".

## B. Matching Human and Model Observer Performance

Models often achieve higher accuracy than human observers. Comparisons across models and human for the same signal contrast can result in ceiling performance for models. To facilitate comparisons, we selected signal peak contrasts for the models that matched average human performance at specific conditions. This allowed us to compare the relative performance across the conditions of interest for models and humans. Figure 4 illustrates the procedure to select contrasts for the ideal observer to match human performance measured in specific conditions in the psychophysical tasks. Similar methods were applied to other model observers to obtain their corresponding signal peak contrast to match human performance.

## C. 3D LKE detection vs. 3D Search: Reversal of rank of order of detectability of signals

We compared human and model observer performance for the microcalcification and mass signals for 3D location known exactly detection and 3D search. In particular, we asked whether the relative detectability of the two signals in the 3D LKE detection task was consistent with 3D search.

Figure 5 shows a dissociation in the average performance (PC) for human observers for microcalcification and mass signals across tasks. For the 3D LKE task, human performance detecting the microcalcification was vastly higher ($\Delta PC = 0.32$) than detecting the mass ($p < 0.01$). Yet, for 3D search performance, this difference was not significantly different across both signals ($p = 0.34$). Figure 5 also shows performance for three typical model observers (IO, NPWE, CHO). To facilitate comparisons of model and human performance, we chose separately for each task (3D LKE detection and 3D search) the model contrast that matched human mass performance. We then used the same contrast for the microcalcification (Figure 4).

Both IO and CHO model observers show a relatively higher performance for the microcalcification than for the mass in both LKE detection and search. Thus, these models predict human performance for 3D LKE detection but fail to predict the deterioration of the microcalcification accuracy for humans in the 3D search. The NPWE model predicts lower detectability for the microcalcification than the mass across both tasks and, thus, also fails to predict human performance.

Finally, Figure 5 also shows the prediction of the FSM model for both tasks. FSM-human utilizes the measured fixations for each observer on each trial, whereas FSM-MAP uses the maximum a posteriori probability algorithm to guide saccades. The FSM model uses a total of 10 templates and also considers multiple eye movements. Its computation thus requires ~80 more times the number of dot products than a regular scanning observer. Due to these computational constraints, the FSM model was computed only using the same signal peak contrast as human observers. For 3D detection, the model predicted higher performance for the microcalcification. For the 3D search, the model predicted higher performance for the mass, although the difference across signal types is larger than that observed in humans. These results were similar for the FSM-human and FSM-MAP.

## D. 2D vs. 3D Search: Interaction between 3D search and signal type

To facilitate human and model comparisons that focused on the effect of 3D search on detectability while avoiding ceiling or chance performance, we selected signal-peak contrasts for the models that matched human 3D performance. This process was done separately for the two signals. The signal peak contrast for each signal was then used for the model observer in the corresponding 2D search task (Figure 4). Similarly to the previous experiment, the FSM model used the same signal peak contrast as human observers.

Figure 6 shows the proportion correct for human observers and model observers for the 2D and 3D search tasks for the microcalcification and mass signals. All standard model observers predicted higher performance in 3D than 2D search for both signal types. In contrast to model observers, human performance also improved in 3D search (vs. 2D search) for the mass but deteriorated significantly for 3D search for the microcalcification signal. The FSM model showed a relative order between 2D and 3D search performance for microcalcification and mass signals similar to that of humans.

To understand the error types, we divided the missed signal trials between search and recognition errors [89]. Search errors are defined as trials in which the observer did not fixate the signal and missed it. Recognition errors are defined as trials in which the observer fixated the signal but still missed it. Figure 7 shows error types for humans and the FSM using human eye movements. The microcalcification signal in 3D was mostly missed for human observers due to search errors, with very few recognition errors in 2D. For the mass signal, search and recognition errors were divided almost evenly in 2D search, with a higher prevalence of search errors in 3D. For the FSM, both search and recognition errors showed a similar pattern to that observed in humans.

## IV. Discussion

Our main finding was the inability of standard model observers to predict the effect of 3D search on the human perceptual performance of small signals (e.g. microcalcifications). These models failed to predict the detrimental effects of 3D search on observers' detection of the microcalcification. For the ideal observer and Channelized Hotelling models, the performance benefited from integrating additional signal slices of 3D (relative to 2D) and outweighed the detriments from the increased location uncertainty of the 3D search space. For these models, there was no detrimental effect in performance specific to searching for small signals in volumetric images. The NPWE model was an exception that resulted in poor performance detecting the microcalcification signal in both 2D and 3D tasks. The NPWE's low detectability for the microcalcification signal arose from the attenuation of the signal's high frequencies by the eye filter. Therefore, the NPWE also failed to predict the interaction between tasks and signal type for human performance.

Analysis of the error types of human observers showed a large increase of 3D search trials in which observers fail to fixate the microcalcification and then erroneously decide that the signal was absent (search errors). This suggests that the deterioration in 3D performance for the microcalcification signal was related to a human tendency not to exhaustively foveate all regions of all slices in 3D search or detect the microcalcification in the visual periphery (Figure 3).

Standard model observers only reflect foveal vision and cannot capture human search process with a visual system that sees the image with varying spatial resolution across the visual field. The proposed foveated search model (FSM) combines elements from model observers from medical image quality [1]–[4], [79], and new components from models of visual search used in vision science [74], [81], [84]. The FSM model can successfully predict the interaction between 3D search and signal type, and the increase in search errors for the microcalcification signal. We evaluated two versions of the FSM: the first version used the actual measured fixations and scrolls from human observers for each trial and applied them to the same images presented to the participants; the second version used an automated saccade generating algorithm that sequentially fixated the most likely signal location (MAP). Although they over-predicted the difference in performance between microcalcification and mass signals for 3D search, both FSM models predicted the main dissociation. This is likely due to a better ability of the FSM relative to humans to integrate information across slices for the mass signal.

Perhaps, the main drawback of the FSM is the computational complexity that makes it non-trivial to implement and time-consuming for iterative optimizations or parameter fitting. The Gabor channels' scaling parameters were fit to a separate data set measuring detectability as a function of retinal eccentricity for both signals. The FSM's computational complexity motivates the need for potential shortcuts to estimating its performance of foveated search models.

There are limitations to the current study. We utilized simulated signals and backgrounds, and trained-observers instead of radiologists. This approach allows for a full knowledge of the statistics of the backgrounds, a pre-requisite for the formulation of the ideal observer, and has historically been the approach for the first step in the development of new model observers [1], [2], [90], [91]. A subsequent step is to extend the results and modeling efforts to real backgrounds with simulated signals or digital phantoms. Preliminary results with digital breast phantoms and radiologists show that the steep deterioration in detectability of small signals in 3D search generalizes to these more realistic backgrounds [92].

## V. Conclusion

The present human and model observer results have several important implications for medical image quality. First, human observer performance in a simple 3D location known exactly detection task does not always reflect performance in more clinically realistic 3D search tasks. Second, search with 3D images introduces an additional source of inefficiency for human performance related to peripheral processing and eye movement exploration. The proposed foveated model observer predicted the relative detectability between the simulated microcalcification and mass in different 3D visual search tasks. Thus, not incorporating these sources of inefficiency in model observers risks mispredicting the impact of new 3D technologies on human observer performance.

# Supplementary Materials

**Stimuli Generation**

Supplementary Supplementary Fig. *1* shows a sample of several consecutive slices of both signals. Signals were designed to contain different ranges of spatial frequencies. While both the background and the MASS signal contain high power in the low frequencies, the MCALC signal contained higher spatial frequencies due to its sharp edges.

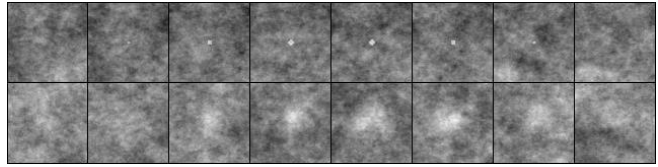

Supplementary Fig. 1: Example of several consecutive slices of the region of interest around a 3D simulated microcalcification (top) and mass (bottom) embedded in noise samples.

**Channelized Hotelling Channel Details:**

Below we describe the details for the channel models used in the current paper.

*Gabor Channels*

We used a set of 1-octave Gabor filters with 8 orientations and 6 central spatial frequencies (0.5, 1, 2, 4, 8, and 16 cycles per degree of visual angle) [69], [86]. Gabor channels are built using the following equation:

$$G_{x,y} = \exp\left(\frac{-4 \ln 2 (x^2 + y^2)}{W_8^2}\right) \cos[2\pi f (x \cos\theta + y \sin\theta)] \tag{S1}$$

where $f$ is the spatial frequency, $\theta$ is the orientation, and $W_8$ is the width.

*Laguerre-Gauss channels*

The channels were generated using the product of Laguerre polynomials and Gaussian functions:

$$u_j(r|a_u) = \frac{\sqrt{2}}{a_u} \exp\left(\frac{-\pi r^2}{a_u}\right) L_j\left(\frac{2\pi r^2}{a_u^2}\right) \tag{9}$$

$$L_j(x) = \sum_{k=0}^{j} (-1)^k \binom{j}{k} \frac{x^k}{k!} \tag{10}$$

using $j = [0, 3, 9, 17]$ and $a_u = [5, 10, 20, 40]$ for a total of 16 channels.

*Difference of Gaussian Channels*

The channels were generated by taking the subtraction of two Gaussian functions:

$$D_n(r) = \exp\left(\frac{-r^2}{Q 2\sigma_n^2}\right) - \exp\left(\frac{-r^2}{2\sigma_n^2}\right) \tag{11}$$

$$\sigma_n = \sigma_0 * \alpha^n \tag{12}$$

using $\sigma_0 = 0.005$, $\alpha = 1.4$, $Q = 1.67$, and $n = [1 \dots 10]$.